# Microscopic Evolution of Hybrid Laboratory Earthquakes

H.O. Ghaffari [1(A)], W.A. Griffith[1], P.M. Benson [2]

1 University of Texas, Arlington, TX, 76019, USA
2 Rock Mechanics Laboratory, University of Portsmouth, Portsmouth, UK

**Abstract**

Characterizing the interaction between water and microscopic defects is one of the long-standing challenges in understanding a broad range of cracking processes. Different physical aspects of microscopic events, driven or influenced by water, have been extensively discussed in numerical calculations but have not been accessible in micro-scale experiments. Through the analysis of the emitted ultrasound excitations during the evolution of individual dynamic microcracking events, we show that the onset of a secondary instability – known as hybrid events in coda part of the recorded waveforms - occurs during the fast equilibration phase of the system, which leads to (local) sudden increase of pore water pressure in the process zone. As a result of this squeezing-like process, a secondary induced instability akin to the long period event occurs. This mechanism is consistent with observations of hybrid earthquakes found in volcanic settings.

**Introduction**

Critical challenges remain in the study of dynamic interactions between in-situ liquids and moving defects (fractures and micro-defects such as dislocations and other topological defects) and in triggering the nucleation and/or movement of defects [1-3]. Such interactions might emit broadband phononic excitations which mirror the complexity of the source dynamics from which they are derived as well as the environment in which the sources propagate.

An important manifestation of these excitations in the geosciences is the study of the seismic response of rocks in the presence of pore fluids. Seismological observations of earthquakes associated with active volcanism have exposed a wide variety of physical phenomena that are manifested as seismic activity [4-5]. In particular, Low-Frequency (LF) earthquakes have been associated with so-called slow slip events in subduction zones and as a consequence of fluid movement during volcanic unrest [6-7]. LF events also known as long-period (i.e., dominant low frequency component in the energy spectrum) and very long-period events, are observed on all types of active volcanoes, often in swarms preceding eruption. Such LF events differ from Volcanotectonic seismicity (VT events) in terms of both their characteristic frequency range and extended harmonic (coda) signature [4,6-7] and have been postulated to be generated from fluid flow and resonance in fractures and conduits within the edifice. Finally, the third type of seismicity shows features of both HF seismicity and also LF harmonic tremor. Known as hybrid events, this type of seismicity is characterized by a high frequency, VT-like onset and a LF-like coda, suggesting that hybrid generation is stimulated by stress regimes leading to both rock failure, and also where fluids are present in order to generate LF and tremor [4,5-8].

Laboratory manifestations of LF, VT, and hybrid seismicity are accessible by recording the Acoustic (phonon) Emissions (AEs) – the laboratory analogue of seismic events in Earth's crust and a commonly-used proxy in laboratory rock physics [9]. The high resolution and sensitivity of sensors used to record AEs allows for detailed study of crack tip evolution over durations of less than tens of nano-seconds at spatial scales as small as the vicinity of the crack tip. In particular, the robust nature of collective excitation modes (such as ultrasound emissions) very close to the source and prior to effects of much fast scattering process is the major advantage in studying dynamics of the system. Benson et al. [9-10] successfully recorded regular, LF, and hybrid events through state-of-the-art experiments in which acoustic excitation signals were recorded by ultrasound sensors placed on basalt rock samples during triaxial deformation experiments under dry and wet conditions.

In this study, by analysing multi-array ultrasound excitations of hybrid events and comparing with their dry counterparts, we elucidate the detailed dynamics of fluid-solid micro-interaction during these experiments. We show that hybrid events are a combination of two types of cracking modes that are generated through either a "crack-like" and "pulse-like" excitations. The key difference between the two is that the pulse-like rupture is self-healing, resulting in a caterpillar or wrinkle-like rupture movement [11-15]. We illustrate how an intermediate phase (in analogy with a "fast healing" process) in the VT section of emitted waves plays a crucial role in inducing a secondary weakening perturbation. The secondary perturbation is accompanied by a rapid decrease in pore pressure, resulting in locking of the micro-fault. Such a mechanism has been hypothesised by Lykotrafitis et al. [12] and briefly by Marone & Richardson [13]. Furthermore, by mapping multi-array recorded phononic excitations on to a spin-like system for a given cracking event, we present a counterpart of the evolution of this intricate system in the context of ordered-disordered transitions. Using this strategy, for the first time, we discover that an oscillatory microscopic squeezing of liquid-layers imprints relatively weaker and longer disturbances of the elastic field; a key characteristic of the LF events.

We use microscopic cracking excitations recorded by multiple sensors during the rock deformation tests. The data sets consist of two rock deformation experiments performed on samples of Mt. Etna basalt under both water- (i.e., wet), and gas- (i.e., dry) saturated conditions (Fig1.a). Both data sets were generated via conventional triaxial deformation experiments, with a suite of 12 sensors arrayed around the test samples to detect AE events. We utilize data from two samples; a water saturated sample with a pore fluid (distilled water) pressure (Pp) of 20 MPa, and a dry sample using Nitrogen gas pressurized at 10 MPa (room temperature). In both cases, a servo-controlled intensifier apparatus maintained an effective pressure of 40 MPa. The samples in both experiments were deformed at a constant axial strain rate of $5 \times 10^{-6}$ s$^{-1}$, controlled via linear variable displacement transducers. To analyze the recorded multi-array acoustic signals, in addition to frequency analysis, we use an innovative dynamic network-based algorithm which maps the sensors and their interactions for each recorded point on nodes and calculates the strength of similarity between nodes as "links" (Methods). We employ this class of time-series analysis together with a previously developed algorithm for multi-array records of AEs (see Methods for details of the algorithm and other techniques). The AE time series recorded at each sensor is represented as a node, and the connection between nodes (i.e., links) in each given time step is then quantified by calculating the similarity measure of the amplitudes. Then, the state of the system (complexity of source, scattering with environment and instrument response) in each time-step is mapped to a network structure where the number of nodes (defining the size of the system) is held constant and the structure of the obtained network is controlled by links.

**Results**

Any valid metric to evaluate an emitted phononic excitation must carry the information regarding both the source and the media through which the phonons travel. For any network, a collection of metrics can be used to characterize the state of the network, and these metrics can provide information about the evolution of both the source and the medium. In this paper, we focus on two main parameters: (1), the temporal evolution of the average number of links to any one node (also referred to as the "degree" of the node), and (2), the "modularity", an index of the community structure of the network as the "bulk" measure (see Methods). The first intriguing result of the analysis is that we find that ultrasound excitations possess patterns of temporal evolution of network parameters that are universal among all recorded events [16-18]. The appearance of the universal patterns in any measure of excited signals shows the robustness of the (collective) process in the source(s) against the much faster scattering processes. The network modularity (Q), a parameter that we describe as a "*Q-profile*" when plotted vs. time, is a measure of the temporal evolution of acoustic networks. For typical excitations, this evolution can be broken down into four main phases (Fig. 1b-c ; Fig.S.2). We refer to the first short phase as the (1) S-phase: an initial apparent strengthening phase; (2) the W-phase is a fast-slip or weakening phase; the fast weakening phase (W) leading to a *peaklike* signal is followed by (3) a sudden declining, i.e., re-strengthening phase, (RS) with a characteristic duration of ~20±3μs for typical dry emissions; and finally (4) the D-phase: a slow slip phase [16-19]. The origin of the RS phase has been postulated to be related to lattice cooling [16, 20] which is the result of equilibration with the environment. The long-term evolution of the D-phase is strongly coupled with scattering properties of the medium (here rock mass) and the magnitude of the source (which is encoded in peak of Q) but short term evolution of this phase remains another issue to be solved. At each excitation level and for short term evolution, $\delta Q$ asymptotically approaches a constant value. For some of the events, $\delta Q$ shows a slow increase in the D-phase for t>180μs (Fig.1c).

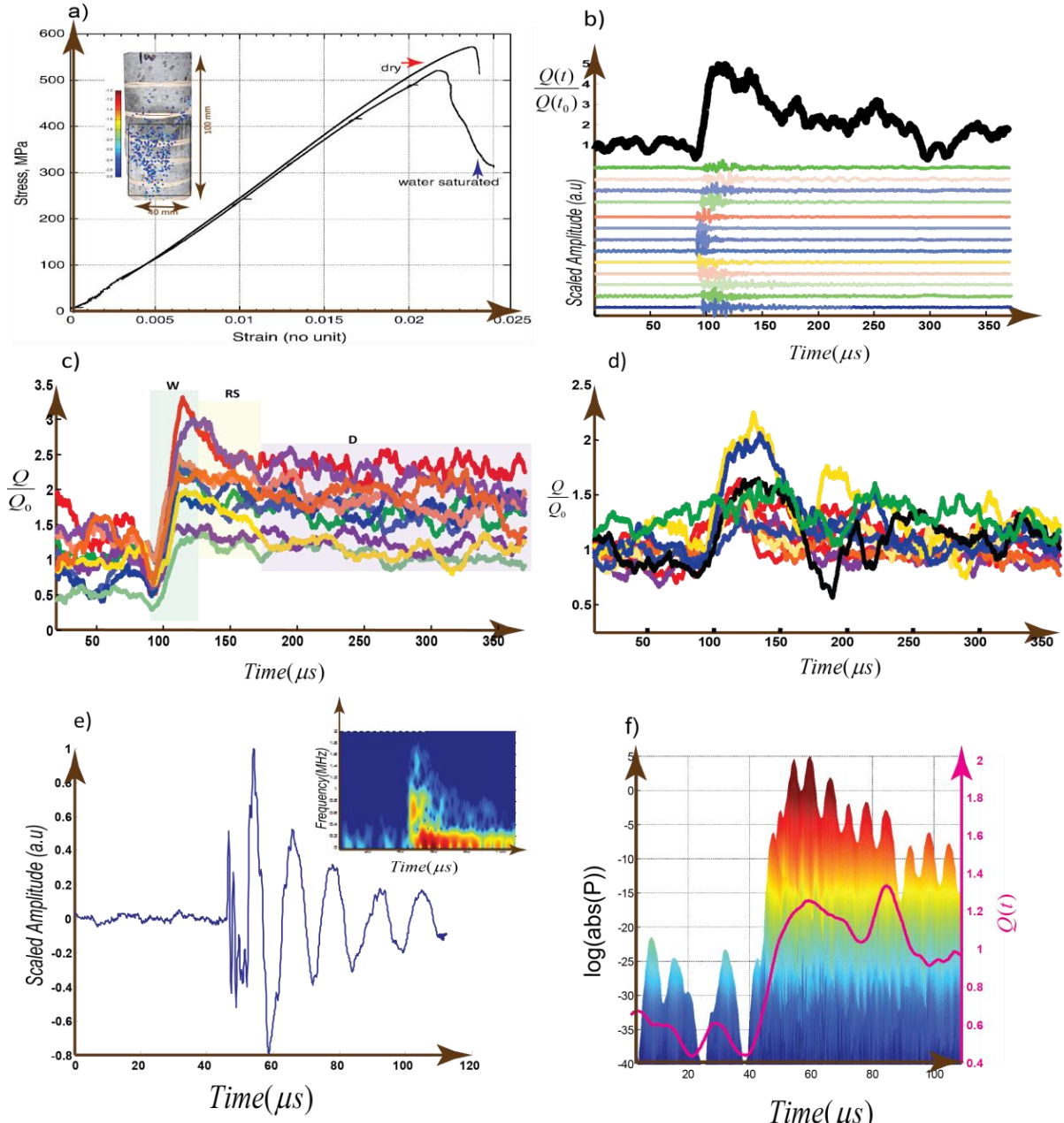

**Figure 1| Study of Q-profiles from microscopic dry and wet events.** (a) The global strain-stress curves for both samples. The inset shows the water saturated sample, with AE locations with sources colored according to maximum recorded amplitude [9] (b) Evolution of Q-parameter and corresponding raw waveforms from a wet-cracking excitation for Event # 42.5 (c) Typical acoustic emission events from dry basalt experiment. The fast weakening phase (W) leading to an impulsive signal is followed by a sudden declining, i.e., re-strengthening phase, (RS) with a characteristic duration of ~20±3μs for typical dry emissions. At each excitation level and for short term evolution, $\delta Q$ asymptotically approaches a constant value. For some of the events, $\delta Q$ shows a slow increase during the D-phase for $t > 180$μs. (d) 10 normalized Q-profiles from water-saturated basalt experiment. Smoother shape – approaching a symmetric shape - of signals is a universal characteristic of wet events. (e, f) A hybrid event with associated spectrogram and Q(t). The inset shows frequency-time plot. $P$ is the Power Spectral Density (PSD) of the waveform and onset of W-phase coincides with the broadening of the power spectrum (Fig.S3). An overlapped (80%) 2,048-point fast Fourier transform is used to calculate the power spectral density.

In general, the aforementioned phases are typical for dry-crackling events and simple friction tests conducted under dry conditions [16-20]-Fig.S.3. Further analysis of acoustic networks shows an intriguing "*pairing*" mechanism, in which k*ink-pair* like structures form in the network parameters, governing the oscillations of Q(t) or other relevant parameters [18]. Furthermore, comparing the trend of Q(t) with the power spectral density (Fig.1e-Fig.S3) shows that the onset of W-phase coincides with the dramatic shift in the dominant frequency. This observation also confirms our previous results that the Q(t) resembles the form of the slip profiles and the stresses measured within granite blocks during propagation of rupture fronts (Fig.S.4).

In contrast with many typical dry-cracking events in which the rate of weakening during the *W*-phase is higher than re-strengthening during the RS-phase, the recorded events under saturated conditions imprint a significantly different shape from their counterpart dry events (Fig. 1d and Fig.2): For wet events the re-strengthening rate during the *RS*-phase is roughly equal to the rate of change during the W-regime, and the net-change of Q(t) approaches zero in a very short time. This is analogous to locking the crack and generating a pulse-like rupture [11-12,14-15]. In Figure 2a we show the corresponding *Q*(*t*) of example hybrid-wet waveforms. In contrast with the dominant evolution of dry events, the RS phase is truncated by another phase, reversing the declining trend of the Q-profile which shortens the duration of the RS phase. As a result, a secondary weakening (double-weakening, DW) phase is superposed onto regular profiles. The DW phase is followed by a rapid decline (L-phase) where Q(t) approaches its initial value prior to the perturbation (Fig.2a). The rapid decline (rapid re-strengthening) is reminiscent of pulse like cracks [12-13]; however, as we will see the mechanism controlling this pulse-like cracking is unique to hybrid earthquakes. The overall shape of Q-profiles of wet events tends to be almost symmetric or with a right-handed asymmetry (negative skewness). To quantify the degree of asymmetry of pulse shapes, we therefore have calculated the degree of skewness [21]. The average skewness of Q-profiles as an asymmetry measure can be quantified by:

$$S = \frac{\frac{1}{t_{max.}} \int_0^{t_{max.}} Q(t)(t-\hat{t})^3 dt}{[\frac{1}{t_{max.}} \int_0^{t_{max.}} Q(t)(t-\hat{t})^2 dt]^{3/2}}, \qquad (1)$$

in which $\hat{t} = \frac{1}{t_{max.}} \int_0^{t_{max.}} Q(t) t \, dt$ is the modularity-weighed mean and $t_{max.} \approx 400 \mu s$. As shown in Figure 2c, most of the wet events are characterized by negative skewness in the Q(t) profile, in agreement with the qualitative observations of asymmetry in Figure 2b and supplementary Figure S.5.

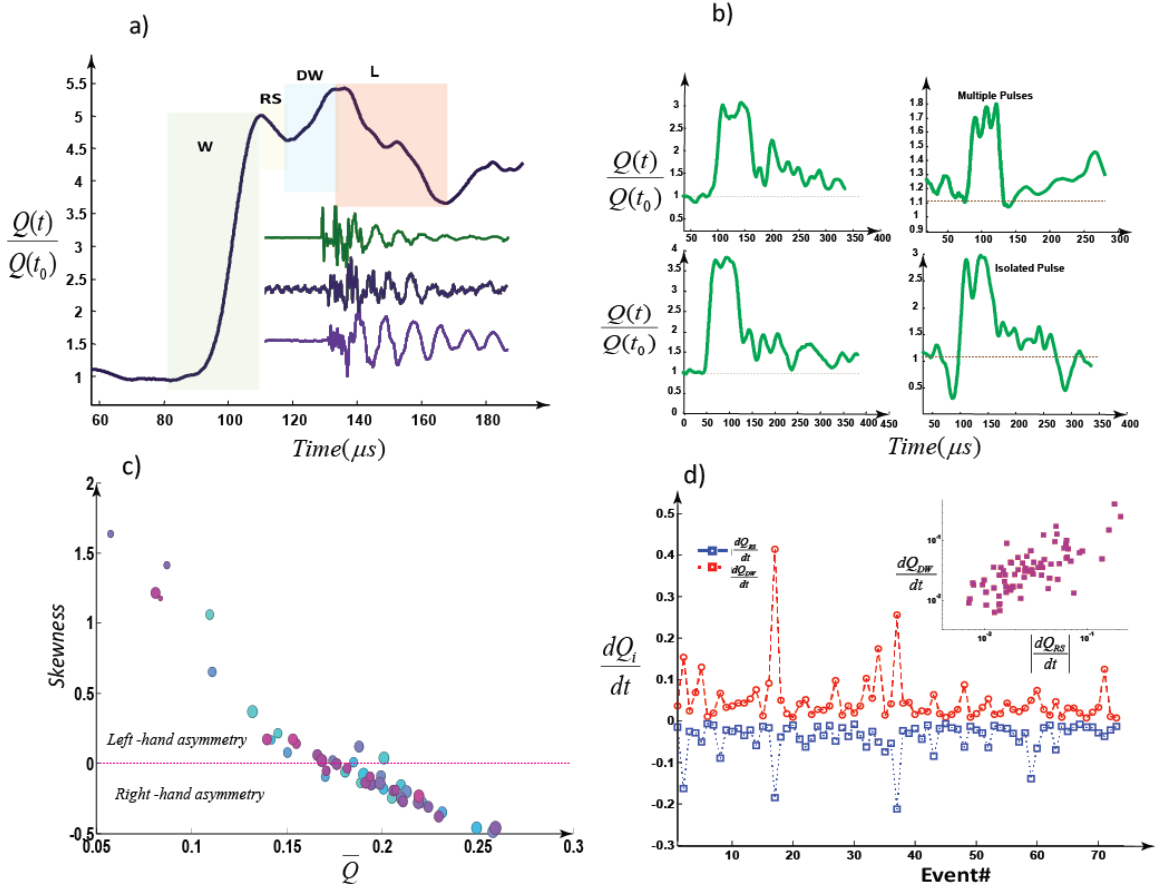

**Figure 2| Examples of acoustic emission events with the signature of the secondary instabilities.** (a) A typical wet-cracking noise and corresponding recorded waveforms. RS-regime is shorter than dry cracking phase and the rate of DW is significantly slower than W-regime. "L" regime is faster-than RS rate and might lead to locking the interface (see more examples in panel (b)-Supplementary Figure S.5). (b) Double or multiple-weakening in wet-cracking noises. The dotted lines are the rest value (normalized) of the Q(t). (c) An approximated classification of 50 wet events using skewness versus temporal-mean of Q(t). Events with right-hand asymmetry are shorter and resemble microscopic pulse-like ruptures. $\overline{Q}$ is temporal average of Q(t) over a $400\mu s$. The circles sizes are proportional to the peak of Q per event (d) The rate of the re-strengthening regime controls the rate of secondary weakening. Events with faster RS-regime induce faster secondary weakening, well described by a power-law relation: $\frac{dQ_{DW}}{dt} \propto \left|\frac{dQ_{RS}}{dt}\right|^{0.8}$. We have shown 70 typical wet –basalt acoustic emission events (table **S1**).

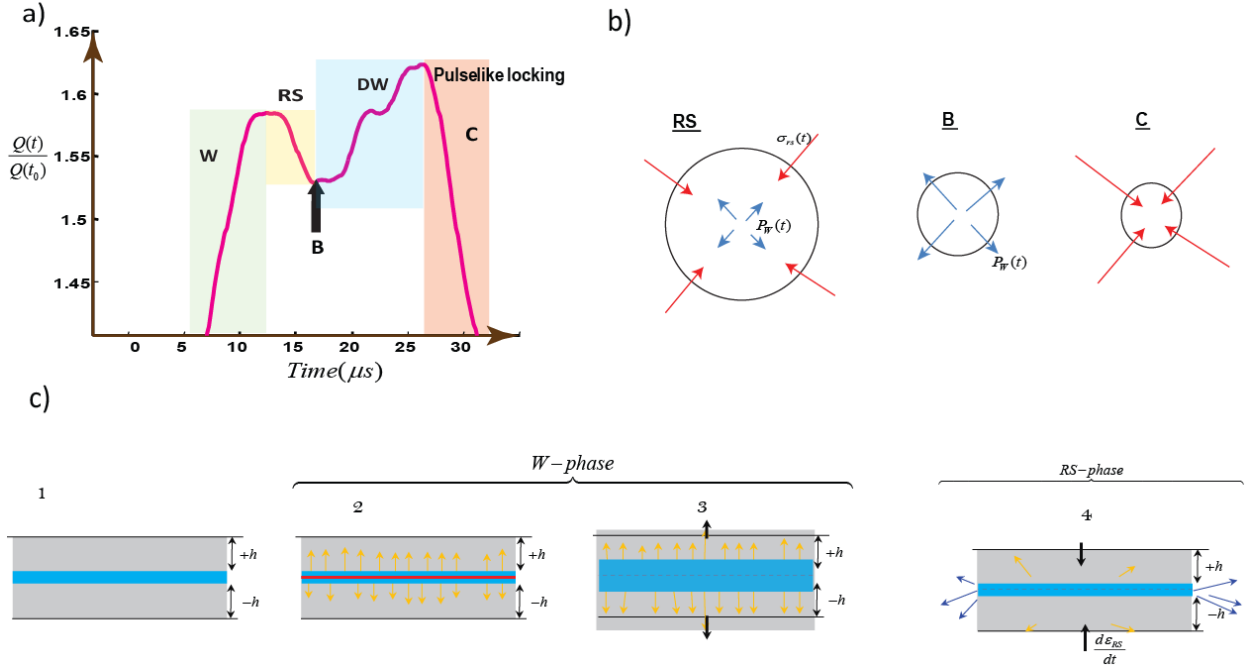

**Figure 3| Double-Weakening Mechanism and fast-release of pore-water pressure.** (a,b) Fast-weakening phase is followed by a re-strengthening phase ("**RS**") which leads to an increase in pore-water pressure in the process zone. This results in a secondary perturbation front ("**B**") with slower rate of weakening than the W-stage (DW). Then, pore pressure, rapidly decreases behind the rupture tip, causing a faster re-strengthening regime and a *locking* of the micro-fault ("**C**"). The final shape of the Q is a *double-impulse* profile whereas single *peaklike* Q(t) characterizes typical dry excitations. (c) Schematic evolution of a "*process zone*". The fracture energy (as the source energy) is deposited within a thin uniform layer of thickness 2h (stage 2). Absorption and diffusion of quasi-particles in this stage results cooling-like process (stage 3). In the re-strengthening (RS) phase, stiffening occurs which yields contraction of water layers and squeezing-out (stage 4).

We propose that the DW and L phases are related to a hydraulic perturbation and rapid release of pore-pressure, and thus fluid movement. To support this idea we examine how the onset of the DW phase becomes separated from the fast-weakening phase (W) by the RS phase, and the rate of DW is strongly correlated with the rate of RS (i.e., the faster the re-strengthening, the faster the double weakening, Fig. 2d). The observed correlation in Fig.2d implies that the source(s) of DW is closely coupled to details of RS evolution. Assuming that RS is the re-loading phase for a process zone which includes a volume of water (Fig. 3b), any change in pore-pressure in the process zone is likely to be associated with the changes in RS-rate and vice versa. Until point B in Figure 3a the source behaves roughly like a dry crack, except for the relatively shorter duration of the RS phase, but at this point the fluid-solid coupling exerts a critical influence on the crack behavior. Considering that the rate of RS is $\sim 10^1$-$10^2$ s$^{-1}$ (i.e., approaching impulsive loading) and water is nearly incompressible, the RS phase results in a rapid increase of water pressure in the process zone (Fig.3c). This process is akin to dynamics of *squeezing-out* the layers of a liquid under pressure [22-23], and we will present further evidence to support this idea. This process results in decrease of volume of water in a series of discreet steps which imprints in the secondary-weakening (onset of DW). Subsequently the onset of decreasing water volume and relating decreasing water-pressure increases the effective stress which causes a faster re-strengthening regime (L-phase) and locking of the micro-fault [12].

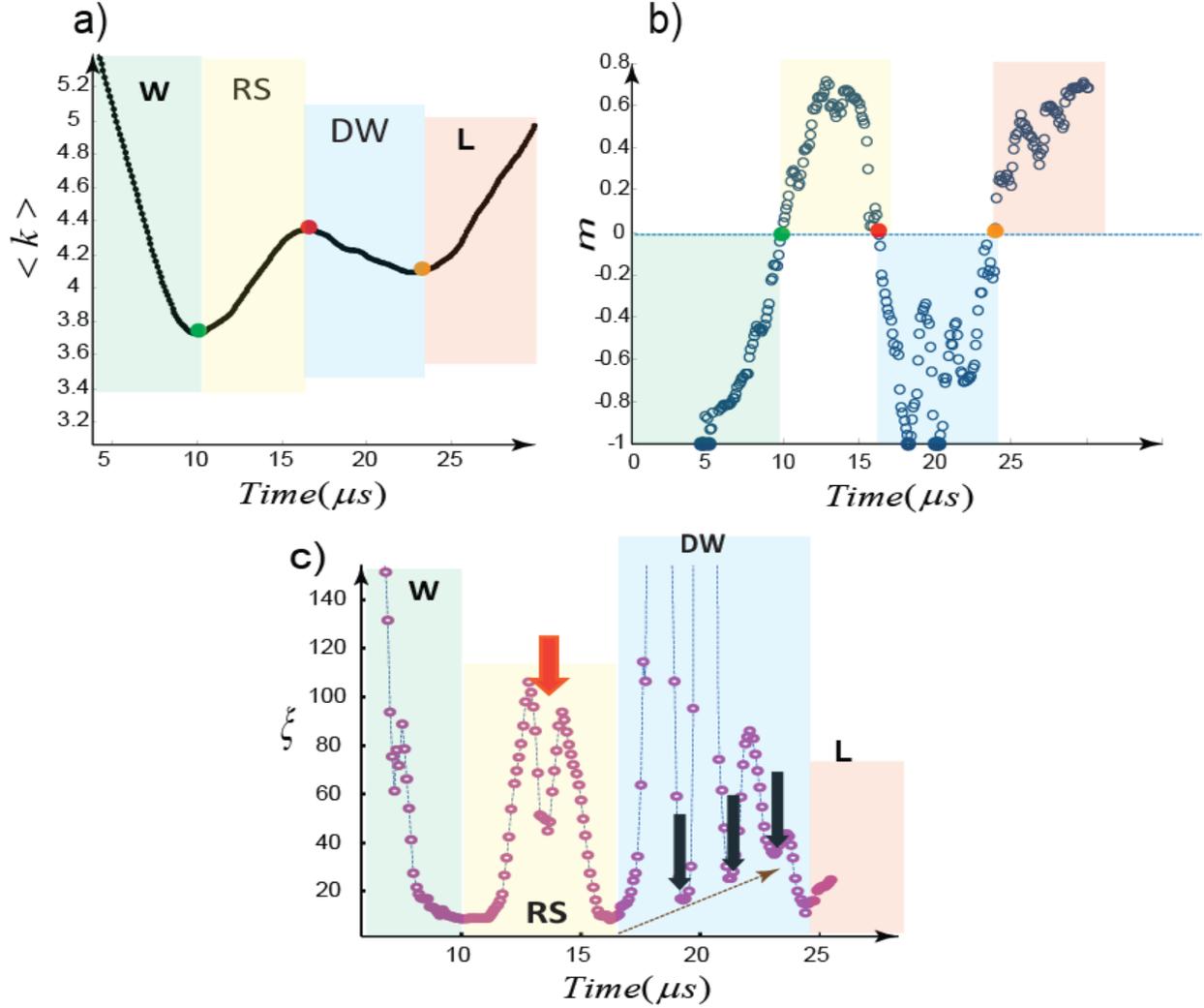

Figure 4| Sequences of W-RS-DW Transitions for a wet-cracking excitation with double-weakening signature (a,b) the average degrees of all nodes (<k>) versus time and the order parameter (*m*) with the critical transition points indicated by color circles. (c) Real time node-node correlation length ($\xi$). The main transitions are indicated by $\xi(t)$ minima. Highly fluctuating order parameter in DW phase drives the system in to the next degenerate state. This signature distinguishes the physics of the secondary weakening phase from the W-phase. We assign this fluctuation to discrete process of squeezing-out of the liquid. The frequency of this almost periodic oscillation is around *200-300kHz*. A double peak in RS phase-exhibited in *m*(t) and $\xi(t)$ - induces sequence of stiffening-softening like feature with durations less than 2µs (The system size –number of nodes-was 300 nodes-Also see Fig.S13-S14).

To explore more features of the double-weakening mechanism, we use the concept of "*K-chains*". K-chains help to visualize the spatiotemporal evolution of acoustic networks and shed light on the microscopic source of Q-profiles [18]. We map the spatial evolution of the degree $k_i$ of the $i^{th}$ node (number of connected links to the node), using polar coordinates $(r_i, \theta_i)_{i=1,...,Nodes}$ where $r_i = k_i$ and $\theta_i$ indicates the position of the node which is fixed on the outer circumference of the cylindrical sample (Fig.S.8-S.11). To analyze the spatiotemporal evolution of *K*-chains, chains are simplified by mapping them as "*pseudo-spins*" where for each node we assign $s_i = sign(\frac{\partial k_i}{\partial t})$ and then $s_i = \pm 1$. With this mapping, each node in a given time-step acquires one of the states (↑ or ↓). Here we use a linear K-chain simplification but the true nature of K-chains are 3d where the state of each node can be represented with a 3d vector such as $\vec{S} \propto \sin\Theta[\cos(\Phi)\hat{x} + \sin(\Phi)\hat{y}] + (\cos\Theta)\hat{z}$ with $\Theta$ as the polar and $\Phi$ as azimuthal angles. In a linear *x*-chain configuration, we have $\Theta = \pi/2$, $\Phi = \pm\pi/2$ and $|\vec{S}| = 1$. We focus on two parameters to characterize the dynamics of chains: (1) the *"order"* parameter $m = <s>$; i.e., mean of up ↑ and down ↓ states; (2) the pair-correlation function (node-node correlation) for the system of nodes. We can fit a correlation function in the form of $G(x) \equiv (1 - \frac{x}{L})\exp(-\frac{x}{\xi})$ where $L$ is the total number of the nodes, *x* is distance, and $\xi$ is the *correlation length*. Correlation length $\xi$ is the cut-off length of the correlation function where for distances shorter than the correlation length, G(x) can be fit by a power law function. For a fully ordered state a triangular function (i.e., fully coherent system and $\xi \to \infty$) is given.

For hybrid events, the RS-phase involves a strong spike in correlation length of the *K*-chains (Fig.4-Fig.S.13-14). A double peak in the RS phase exhibited in *m*(t) and $\xi(t)$ induces a sequence of loading-unloading oscillations, each with a duration less than *2μs*. Another highlighted feature of hybrid events is a highly fluctuating order parameter in the DW phase which differentiates the nature of the secondary weakening phase from that of the W-phase. The secondary weakening phase is characterized by successive returns of the correlation length to local minimums, suggesting rapid crossings of the critical transition points between highly metastable regimes where successive local perturbations drive the system to the next transition point. This discrete sequence increases (on average) the minimum correlation length as the system approaches the locking phase (Fig.4). The frequency of this quasi-periodic oscillation of $\xi$ in the DW regime is around *200-300kHz* for our tests. We ascribe this peculiar dynamic behavior to squeezing-out of the pore water yielding a decrease in the water volume, consistent with results of recent experiments and hydro-dynamic numerical simulations of thin-film liquids [22]. The vibration of the crack in the DW regime can also be modeled with Chouets's crack-vibration formulation which was used to relate volcano-seismicity to fluid flow in fractures and conduits within a volcanic edifice [4-5].

**Discussion**

We quantified cracking excitations, and in the process have developed a new understanding of laboratory AEs by comparing rock deformation experiments in water saturated (wet) and dry conditions. To achieve this, we have developed the concept of *K-chains* to shed more light on micro-evolution of structures in acoustic networks. These structures reflect the processes involved in the vicinity of the propagating microcrack tips. This is crucial, as the moving crack tip generates the source of the AE energy. While we showed that the fast-release of pore-pressure generates hybrid events due to pulse-like cracking, such waveforms have, in certain conditions, also been reported in dry conditions [12, 14]. Furthermore, it has been shown that their occurrence is more probable at lower confinement stresses [15]. The occurrence of self-healing pulses is mostly analyzed in terms of certain conditions on state-and-rate equations and the low-stress range [15]. Reflections of waves from local heterogeneities are also believed to assist cessation of slip behind the pulse [24]. In the context of configurations of K-chains, we can offer a new explanation for the occurrence of the self-healing regime. For typical dry events, the oscillation of slow thermalization part (i.e., short term D-phase prior to strong effect of scattering) and approaching a steady state value with a slight growth trend implies peculiar slow healing processes [20]. This process is imprinted in K-chains as *"healed"*-like sites in the form of proliferated or extended healed-like elements. We believe this is the general processes that occurs in microscopic crack-like signals whereas the suppression of this phase leads to pulse-like cracking. We will investigate further features of such mechanisms in our future studies. Furthermore, the abstract formulation for RS to DW transition (Supplementary Information-Section2) can be extended through the addition of a poro-elastic element, and with hydrodynamic equations, to describe squeeze-out dynamics of water-layers.

The role of capillary bridges has been neglected in this research which adds negative pore-pressure as the capillary bridges add an appreciable drag force [25-26]. These capillary bridges act like liquid bonds, resulting in additional frictional resistance. While we considered a simplified linear fictitious spin chains and analyzed corresponding order parameter oscillations, extension of them to 3D K-chains will provide a unique opportunity to shed more lights on transition between phases. Among them is to study the details of short term oscillation of D-phase, immediately after RS phase, which is believed to represent the slow thermalization dynamics as the unique characteristic of non-equilibrium process. Another interesting study could focus on the dynamics of K-chains in different temperatures as the major parameter in study of volcanic seismicity.

**Materials and Methods**

**Laboratory Procedures:**

Dry and Wet experiments on Basalt-rock samples: Samples of basalt from Mt. Etna volcano (approximately 3.8% porosity) were deformed using a standard triaxial deformation apparatus installed at *University College London* (UK). Cylindrical samples 40 mm in diameter and 100 mm in length were isolated from a confining medium (silicone oil) via an engineered rubber jacket containing inserts for mounting piezoelectric sensors in order to detect AE events-Fig.S1. AE event signals (voltages) are first pre-amplified 40 dB, before being received and digitized.

**Networks of Acoustic emission waveforms:**
To evaluate reordered multiple acoustic emission (multiple time series for an occurred event), we use a previously algorithm on waveforms from our reordered acoustic emissions [16-19]. The main steps of the algorithm are as follows:

(1) The waveforms recorded at each acoustic sensor are normalized to the maximum value of the amplitude in that station.

(2) Each time series is divided according to maximum segmentation, in a way that each segment includes only one data point. The amplitude of the *j*th segment from *i*th time series ($1 \leq i \leq N$) is denoted by $u^{i,j}(t)$ (This is in the unit of m-Volt). *N* is the number of nodes or acoustic sensors. We put the length of each segment as a unit. This considers the high temporal resolution of the system's evolution, smoothing the raw signals with 20-40 time windows (~400-800ns).

(3) $u^{i,j}(t)$ is compared with $u^{k,j}(t)$ to create an edge among the nodes. If $d(u^{i,j}(t), u^{k,j}(t)) \leq \zeta$ (where $\zeta$ is the threshold level discussed in the following point) we set $a_{ik}(j) = 1$ otherwise $a_{ik}(j) = 0$ where $a_{ik}(j)$ is the component of the connectivity matrix and $d(\bullet) = \|u^{i,j}(t) - u^{k,j}(t)\|$ is the employed *similarity metric*. With this metric, we simply compare the amplitude of sensors in the given time-step. The employed norm in our algorithm is absolute-norm.

(4) Threshold level ($\zeta$): To select a threshold level, we use an introduced method in [16-17] where uses an adaptive threshold criterion and is stable for an range of deviations from $\zeta$. The result of this algorithm is adjacency matrix which its components are given by $a(x_i(t), x_k(t)) = \Theta(\zeta - |u^{i,j}(t) - u^{k,j}(t)|)$ where $\Theta(...)$ is the Heaviside function.

In general, the modularity of a network measures the degree of division of that network into modules: if a network has high modularity, the strength of connections in individual modules is strong, whereas the strength of connections between modules is not. The network's modularity characteristic is addressed as the quantity of densely connected nodes relative to a null model (random model). The main diagnostic in this work is the *Q-profile*. The modularity is the result of some optimization of the cluster structure of a given network. The modularity *Q* (i.e., objective function) is defined as [27]:

$$Q = \sum_{s=1}^{N_M} [\frac{l_s}{L} - (\frac{d_s}{2L})^2], \quad (A.1)$$

in which $N_M$ is the number of modules (clusters), $L = \frac{1}{2}\sum_{i}^{N} k_i$, $l_s$ is the number of links in module *s* and $d_s = \sum_{i} k_i^s$ (the sum of node degrees in module *s*).

We use the Louvian algorithm to optimize Eq.A1 [28], which has been used widely to detect communities in different complex networks. Then, in each time step during the evolution of waveforms (here over observation windows of ~400 μs), we obtain a Q value. The temporal evolution of Q values in the monitored time interval forms the Q-profile.

The introduced network algorithm is in close relation with space-correlation methods as well as temporal frequency analysis (Fig.S3 and Fig.S.12). We can define a similar measure to *time-widowed correlation method* [29-30] where the inner product is replaced with Heaviside function. Let us consider a sequence of nodes in a certain time step and the length of this sequence is L=2*l* and the amplitude of each node is $u(x)$. The space windowed correlation is given by[29,31]:

$$R(x_s) = \frac{\int_{x-l}^{x+l} u(x')u(x' + x_s)dx'}{\int_{x-l}^{x+l} u^2(x')dx'} \quad (A.2)$$

where the space-window is centered at length *l* with duration 2*l* and $x_s$ is the space shift used in the cross correlation. Here the employed norm is inner-product and can be replaced with another norm :

$$R(x_s) \propto \int_{x-l}^{x+l} \|u(x'), u(x'+x_s)\| dx' \qquad (A.3)$$

Summing over all space-shifts and replacing the norm with a similarity metric we get:

$$\rho \propto \sum_{x_s} \sum_{x'} \Theta(\zeta - \|u(x') - u(x'+x_s)\|)$$

(A.4)

Then, $\rho$ is proportional to the density of links owning to the employed metric where we use to construct the links between nodes represented by pairs of $u(x'), u(x'+x_s)$. To confirm this calculation, we repeated the algorithm with changing the metric to $a(x_i(t), x_k(t)) = \Theta(\zeta - |u^{i,j}(t) \bullet u^{k,j}(t)|)$ with the inner-product norm similar to (A.2). The results show a clear peak-like trend of <k> indicating that the density of links represents space-correlation of sites.

## Supplementary Information

Supplementary information accompanies this paper.

## Acknowledgements

This material is based upon work supported in part by the U. S. Army Research Laboratory and the U. S. Army Research Office under contract/grant number _W911NF1410276.

## Author Contributions

All authors contributed to the analysis the results and reviewed the manuscript. H.O.G. performed the calculations. H.O.G and W.A.G co-wrote the manuscript. P.B designed the main tests and helped interpretations of the results. W.A.G and P.B supervised the research, provided part of the data sets and helped to analyze the results.

## Competing interests statement

The authors declare no competing financial interests.

# Microscopic Evolution of Hybrid Laboratory Earthquakes

H.O. GHAFFARI [1(A)], W.A. GRIFFITH [1], P.M. BENSON [2]

*1 University of Texas, Arlington, TX, 76019, USA*
*2 Rock Mechanics Laboratory, University of Portsmouth, Portsmouth, UK*

*General Points of the Supplementary Information*

In this supplementary information, we present further examples of double (and multiple) perturbations in wet cracking noises.

- The data sets has been recorded in different loading stages of our tests and is not restricted to a certain stage of the loading time.
- The details of the employed method to analysis Acoustic excitations and the acquisition system (ASC system) have been reported in our previous publication(Benson et al 2010 ;Benson et al 2008). The schematic picture of the experimental set-up has been shown in Fig.S.1.
- Duration of RS-phase and hydro-mechanical formulation of RS-DW phases
- K-chains :further examples and mapping near-filed acoustic excitation on spin systems

## 1. *Some general points on experiments*

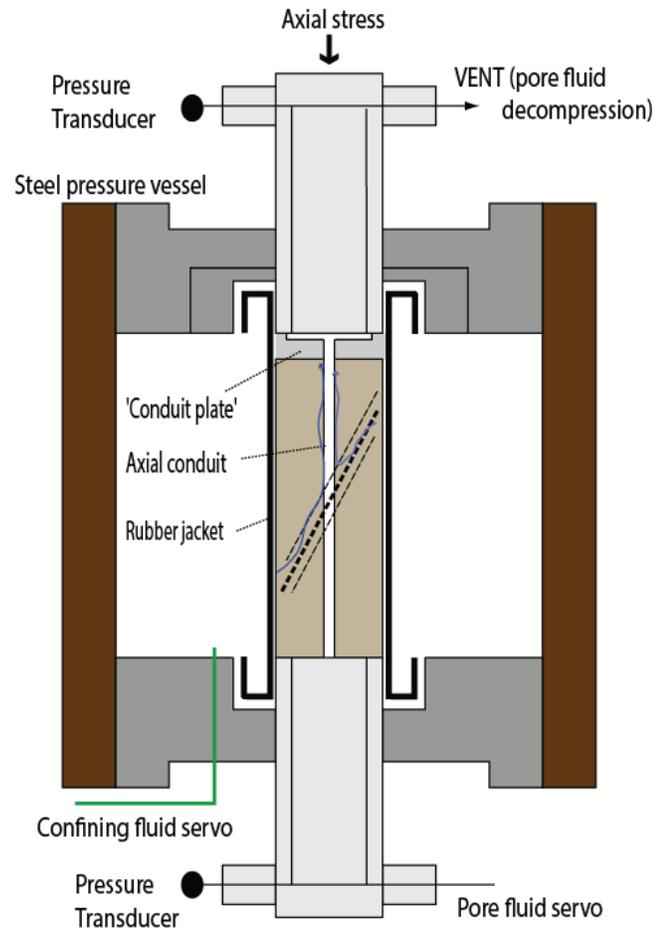

**Figure S.1 A schematic picture of Experimental setup [modified from 2 ]**

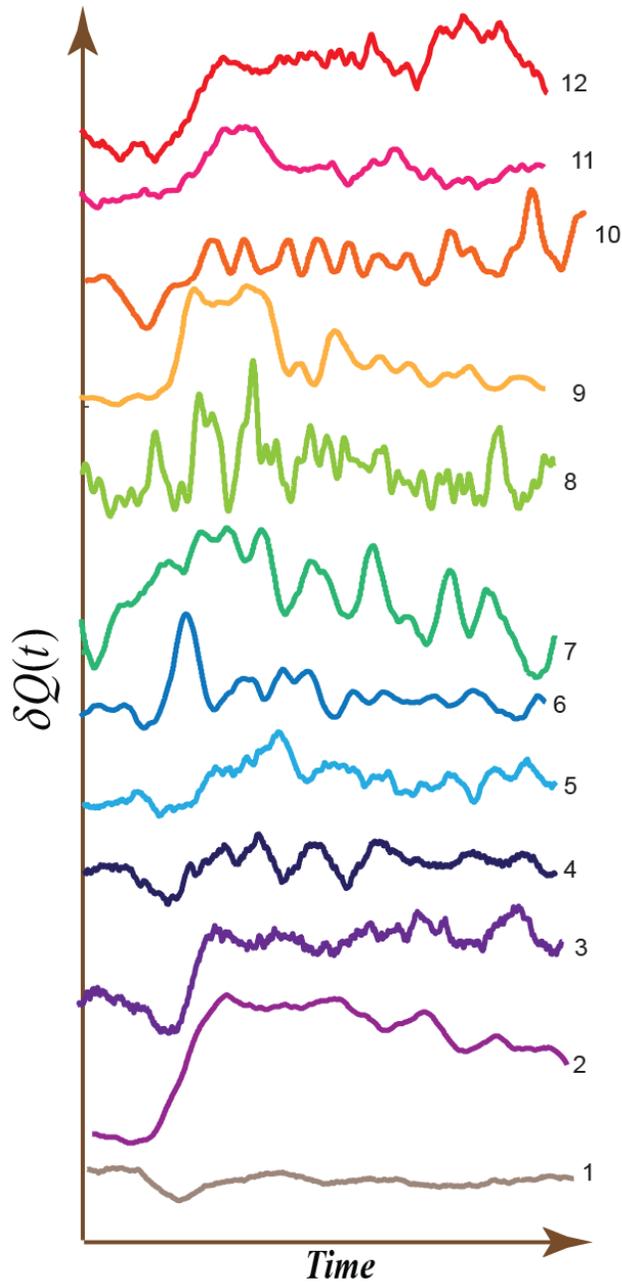

**Figure S.2 Spectrum of Q-profiles from different experiments.** The experiments are as follows: 1-3) Dry Westerly granite experiments in a triaxial cell ;(4-5) Gypsum in room temperature (triaxial test) ;(6) Dry pulse-like crack (7) dry excitation from indentation of a thin mica-film(8)dry multiple-pulse like excitation (9) Double peak-like wet-event (10) tremor-like excitation with train of peaks in tail of Q(t) . A volume of *trapped water* in a thin-plastic sphere is hit by an impulsive loading (in split Hopkinson pressure bar)   (11-12) sandstone cracking noises in a true triaxial test . (Time and Q(t )are not scaled).

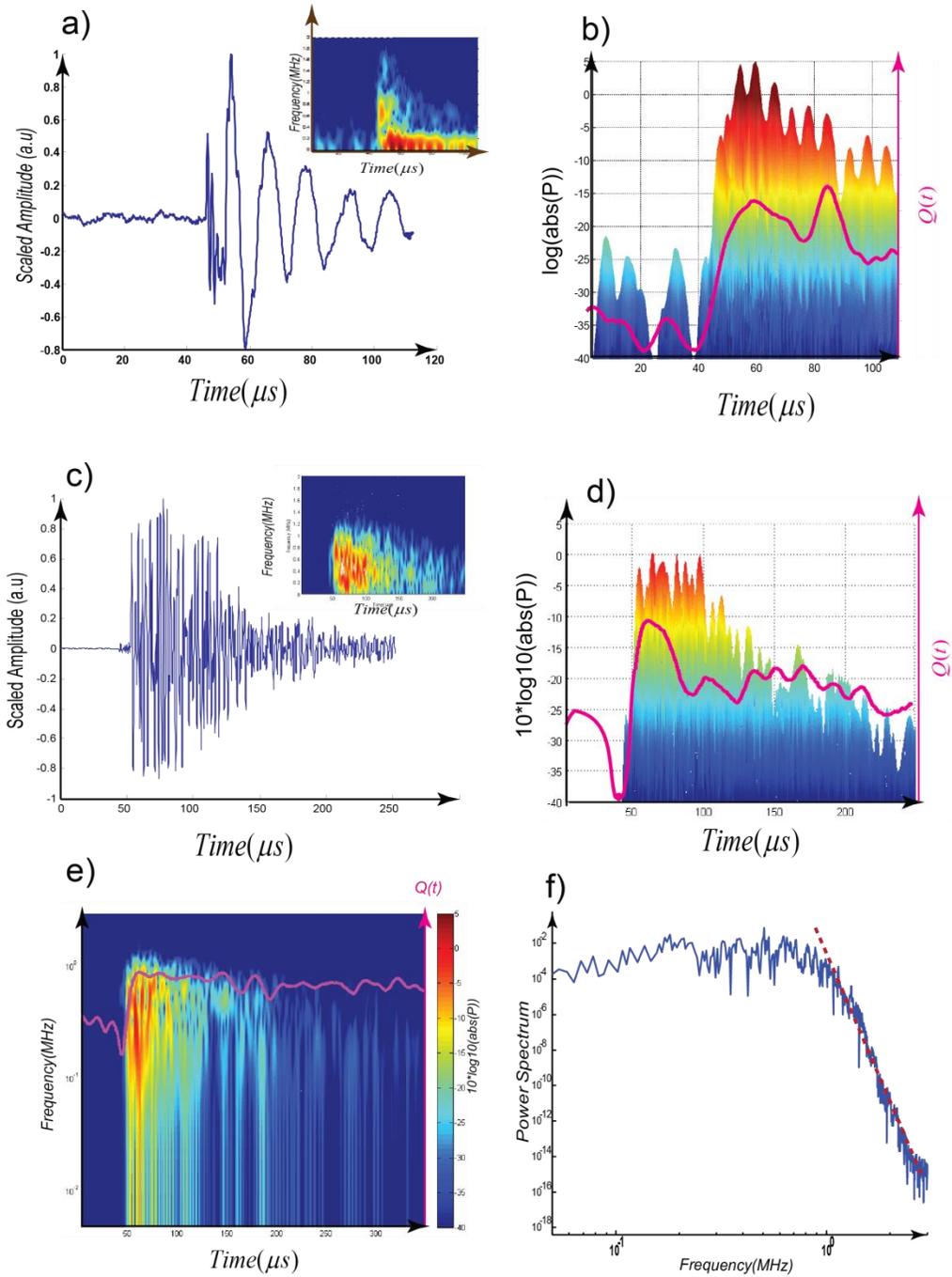

**Figure S.3 Hybrid (a,b) and Dry (c,d) events with associated spectrograms and Q(t). The insets show the spectral amplitudes versus frequency and time**. With this comparison, we find that the onset of high-frequency regime almost matches with the W-phase in Q(t). P is the power spectral density . Onset of W-phase is almost correlated with the broadening of the power spectrum. An overlapped (80%) 2,048-point fast Fourier transform is used to calculate the power spectral density. (e,f) another example from a dry cracking event with the corresponding spectrum.

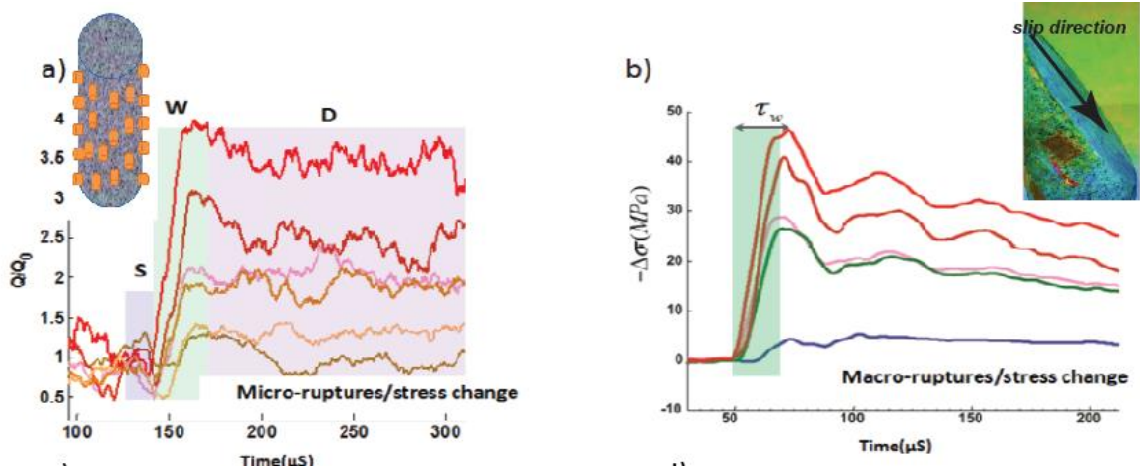

**Figure S.4. (ab)** Q(t) profiles from micro-cracking excitations in dry case and major slip records with strain gauges (dynamic stress field) on saw-cut frictional surfaces of Granite.

## 2. *Characterization of wet-cracking excitations*

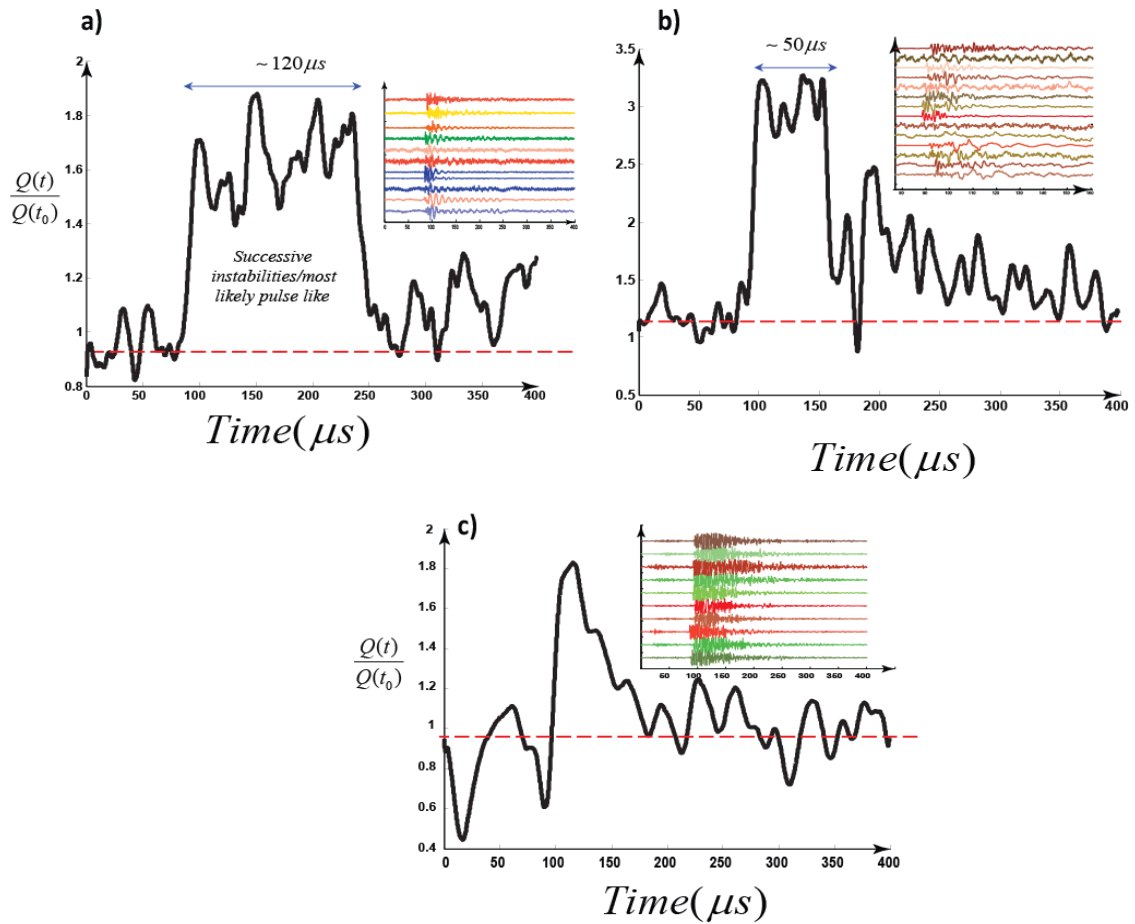

**Figure S.5.** *Pulse-like* **ruptures (slip pulses) are frequent in "wet events". (a,b) Rapid-drop of the Q after the peak point indicates fast re-strengthening phase (i.e., locking zone), dissimilar to typical crack-like ruptures (c).** In panel (a) ,we have shown a hybrid event with a very long duration of secondary instability with successive pulse-like instabilities (multiple peak-like exciations) .

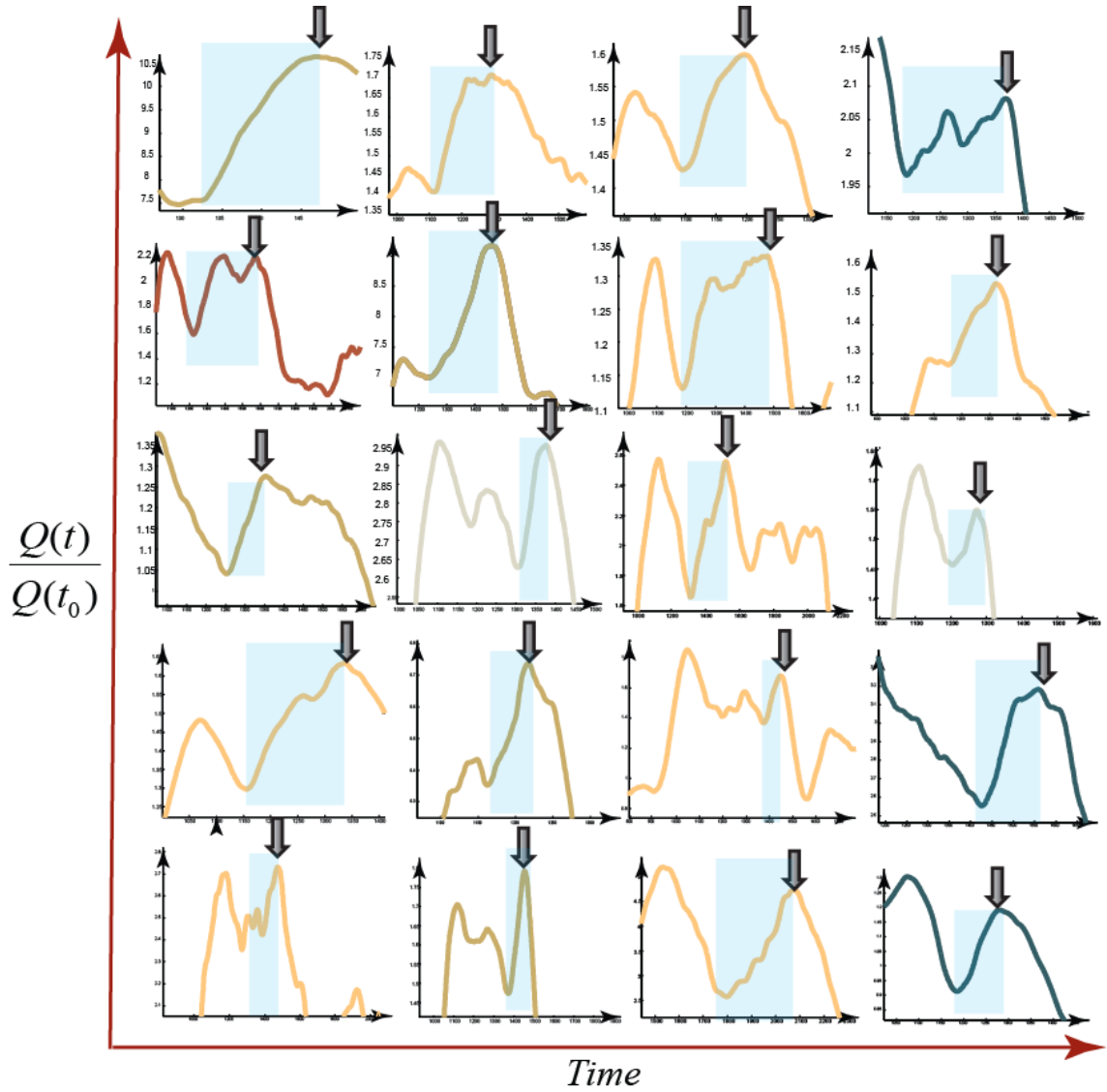

**Figure S.6 Examples of secondary instability in Wet events .The arrow indicates the peak of the secondary instability.**

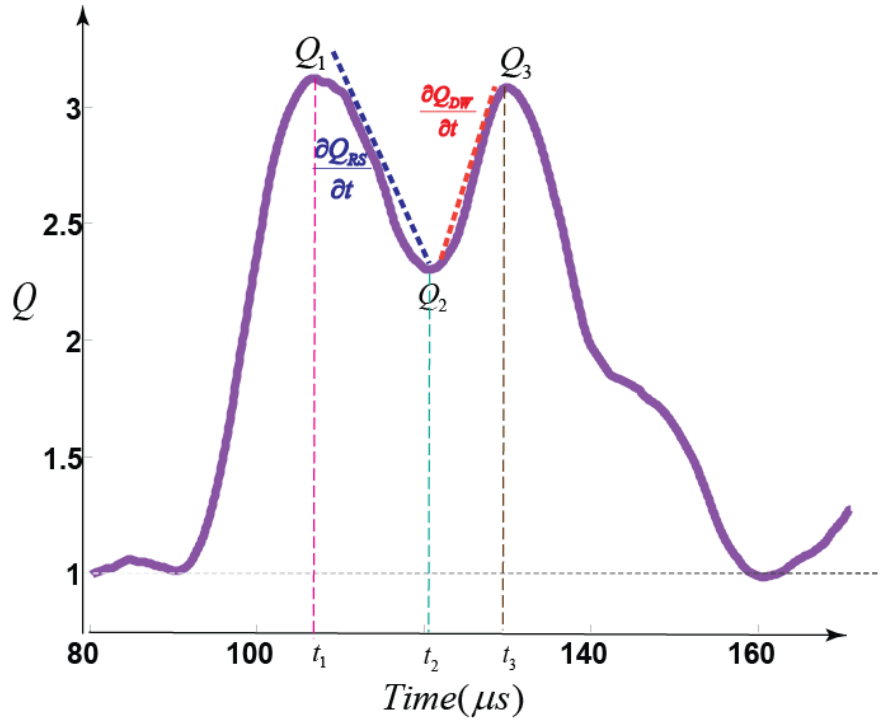

**Figure S.7 A typical Q-profile with double-weakening signature as the result of hybrid events**

| Event# | $t_1$(*10)µs | $Q_1$ | $t_2$(*10)µs | $Q_2$ | $t_3$ (*10)µs | $Q_3$ |
|---|---|---|---|---|---|---|
| 1 | 1156 | 1.878 | 1265 | 1.721 | 1372 | 2.106 |
| 2 | 1118 | 10.55 | 1383 | 6.214 | 1576 | 9.171 |
| 9 | 1164 | 2.143 | 1266 | 1.895 | 1448 | 2.344 |
| 12 | 1139 | 2.157 | 1322 | 1.639 | 1485 | 2.755 |
| 13 | 1042 | 3.5331 | 1073 | 3.376 | 1125 | 4.054 |
| 15 | 1049 | 1.266 | 1063 | 1.256 | 1115 | 1.3115 |
| 17 | 859 | 1.111 | 922 | 1.051 | 1040 | 1.271 |
| 19 | 1085 | 3.004 | 1189 | 2.072 | 1330 | 3.024 |
| 26 | 917 | 2.106 | 1027 | 1.873 | 1150 | 2.277 |
| 28 | 920 | 2.109 | 1029 | 1.837 | 1153 | 2.275 |
| 38 | 1137 | 1.937 | 1210 | 1.806 | 1368 | 2.495 |
| 43 | 1124 | 2.54 | 1213 | 2.228 | 1279 | 2.509 |
| 61 | 1086 | 2.388 | 1145 | 2.264 | 1250 | 2.819 |
| 62 | 1055 | 2.917 | 1191 | 2.105 | 1293 | 2.878 |
| 68 | 1226 | 1.718 | 1325 | 1.582 | 1384 | 1.653 |
| 78 | 878 | 1.135 | 948 | 1.02 | 1030 | 1.77 |
| 81 | 1129 | 18.73 | 1447 | 12.82 | 1572 | 18 |
| 97 | 1129 | 3.146 | 1202 | 2.867 | 1341 | 3.552 |
| 104 | 1058 | 2.318 | 1118 | 2.205 | 1175 | 2.305 |
| 122 | 996 | 1.419 | 1102 | 1.309 | 1294 | 1.473 |
| 125 | 1045 | 2.808 | 1138 | 2.394 | 1239 | 2.805 |
| 126 | 1037 | 2.464 | 1150 | 1.758 | 1458 | 3.346 |
| 138 | 1016 | 2.238 | 1200 | 1.472 | 1478 | 1.917 |
| 139 | 1077 | 3.064 | 1103 | 3.028 | 1158 | 3.176 |
| 147 | 971 | 2.105 | 1059 | 1.788 | 1231 | 2.249 |
| 151 | 1041 | 1.9 | 1124 | 1.799 | 1272 | 2.339 |
| 163 | 1111 | 7.322 | 1191 | 6.935 | 1432 | 9.2999 |
| 187 | 1100 | 1.61 | 1196 | 1.454 | 1265 | 1.55 |
| 193 | 1056 | 2.024 | 1228 | 1.391 | 1336 | 1.78 |
| 195 | 1139 | 1.297 | 1229 | 1.227 | 1292 | 1.343 |
| 197 | 1099 | 2.976 | 1174 | 2.727 | 1216 | 2.879 |
| 213 | 1159 | 6.505 | 1184 | 6.347 | 1243 | 6.953 |
| 219 | 1084 | 2.934 | 1170 | 2.712 | 1216 | 2.965 |
| 224 | 1134 | 2.113 | 1229 | 1.639 | 1290 | 2.702 |
| 235 | 1082 | 3.165 | 1224 | 2.11 | 1662 | 2.702 |
| 246 | 1041 | 3.73 | 1071 | 3.569 | 1109 | 3.729 |
| 261 | 1069 | 6.734 | 1113 | 5.8 | 1162 | 7.051 |
| 264 | 908 | 1.773 | 976 | 1.613 | 1037 | 1.875 |
| 276 | 1137 | 2.294 | 1162 | 2.219 | 1223 | 2.487 |
| 311 | 1059 | 1.48 | 1115 | 1.38 | 1190 | 1.494 |
| 325 | 1073 | 3.822 | 1132 | 3.568 | 1182 | 3.687 |
| 330 | 990 | 1.123 | 1057 | 1.057 | 1197 | 1.371 |
| 335 | 1065 | 5.299 | 1151 | 4.576 | 1325 | 5.673 |
| 338 | 1108 | 1.534 | 1247 | 1.272 | 1339 | 1.426 |
| 356 | 989 | 1.496 | 1076 | 1.435 | 1099 | 1.456 |
| 357 | 1002 | 1.321 | 1096 | 1.158 | 1309 | 1.509 |
| 362 | 997 | 2.492 | 1058 | 2.372 | 1133 | 2.577 |
| 365 | 1053 | 2.456 | 1216 | 1.442 | 1295 | 2.136 |
| 374 | 1300 | 1.421 | 1357 | 1.341 | 1404 | 1.383 |
| 377 | 1044 | 1.429 | 1136 | 1.226 | 1193 | 1.326 |
| 378 | 1068 | 1.601 | 1218 | 1.167 | 1318 | 1.493 |
| 382 | 1047 | 3.564 | 1118 | 3.11 | 1183 | 3.459 |
| 387 | 1061 | 1.041 | 1131 | 0.9551 | 1205 | 1.076 |
| 413 | 1109 | 1.179 | 1211 | 1.015 | 1274 | 1.124 |
| 421 | 1047 | 2.137 | 1081 | 2.075 | 1148 | 2.363 |
| 425 | 1042 | 1.735 | 1109 | 1.536 | 1165 | 1.691 |
| 429 | 1129 | 3.309 | 1260 | 2.635 | 1326 | 2.786 |
| 431 | 1212 | 1.279 | 1393 | 0.776 | 1491 | 1.081 |
| 445 | 1123 | 5.261 | 1306 | 2.707 | 1566 | 4.021 |
| 478 | 1129 | 4.18 | 1224 | 3.561 | 1321 | 4.269 |
| 479 | 1047 | 1.432 | 1164 | 1.247 | 1225 | 1.413 |
| 482 | 1116 | 1.246 | 1222 | 1.137 | 1260 | 1.19 |
| 497 | 1183 | 6.683 | 1280 | 6.014 | 1326 | 6.224 |
| 504 | 1062 | 1.576 | 1133 | 1.478 | 1264 | 1.591 |
| 509 | 1052 | 1.519 | 1127 | 1.334 | 1235 | 1.672 |
| 510 | 1132 | 1.504 | 1298 | 1.251 | 1398 | 1.57 |
| 515 | 897 | 1.454 | 1004 | 1.372 | 1124 | 1.611 |
| 528 | 1107 | 1.361 | 1168 | 1.275 | 1247 | 1.329 |
| 541 | 1046 | 2.782 | 1145 | 2.627 | 1213 | 2.77 |
| 544 | 1023 | 1.721 | 1174 | 1.296 | 1256 | 1.567 |
| 546 | 1111 | 4.126 | 1151 | 3.976 | 1234 | 5.014 |
| 552 | 1073 | 2.689 | 1109 | 2.608 | 1169 | 2.674 |
| 558 | 1064 | 1.371 | 1137 | 1.28 | 1221 | 1.334 |

**Table S1. List of the used Typical hybrid events to plot Fig.11. See the definition of the parameters in Fig.S3.**

To proceed, we estimate the duration of the RS phase during a single hybrid event. The rate of increasing fluid pressure ($P_w(t)$) in the RS regime is proportional to the rate of re-strengthening:

$$\left|\frac{\partial Q_{RS}}{\partial t}\right| \cong k_{RS} \frac{\partial P_w(t)}{\partial t} . \tag{S.1}$$

This statement also is true for the DW regime where at the onset of this phase the pore-pressure is in its maximum value, inducing secondary weakening, and at the end of this phase the pore-pressure approaches zero:

$$\frac{\partial Q_{DW}}{\partial t} \cong -k_{DW} \frac{\partial P_w(t)}{\partial t} . \tag{S.2}$$

The coefficients $k_{RS}$ and $k_{DW}$ indicate the interaction of water and solid body in loading (RS) and unloading regimes (DW), respectively. Furthermore, we estimate empirically (Fig.2d-inset/main text):

$$\frac{dQ_{DW}}{dt} \propto \left|\frac{dQ_{RS}}{dt}\right|^{\beta}, \tag{S.3}$$

where $\beta \approx 0.8$ (Fig.2d-main text). We can assume a <u>linear growth of pore-fluid pressure in the RS regime</u>:

$$\frac{\partial P_w(t)_{RS}}{\partial t} \propto t, \tag{S.4}$$

Plugging (5) in (2) and (3) and using (5), we obtain:

$$P_w(t)_{t \in RS} \propto \left(\frac{k_{RS}}{k_{DW}}\right) \frac{t^{\beta+1}}{\beta+1} \tag{S.5}$$

and then,

$$P_{\max}(w) \propto \left(\frac{k_{RS}}{k_{DW}}\right) \frac{t_{RS}^{\beta+1}}{\beta+1} \tag{S.6}$$

To trigger the secondary perturbation, we should satisfy $P_{\max} \geq \sigma_{th}$ where $\sigma_{th}$ is a stress-threshold quantity for the hydraulic perturbation and (7) is simplified to yield:

$$t_{RS} \geq \Theta \sigma_{th}^{\frac{1}{1+\beta}}, \tag{S.7}$$

with $\Theta = \left(k_{DW} \frac{\beta+1}{k_{RS}}\right)^{\frac{1}{1+\beta}}$. From Figure 11, $\beta \approx 0.8$ allowing a lower bound for the duration of the RS regime to be approximated as: $t_{RS} \geq \Theta \sqrt{\sigma_{th}}$. This indicates that increasing $\sigma_{th}$ will stretch the minimum duration of the RS regime. In contrast, increasing $\beta$ shrinks the duration of $t_{RS}$, leading to earlier onset of the DW phase.

3. *Further characterization of wet-cracking excitations with K-chains*

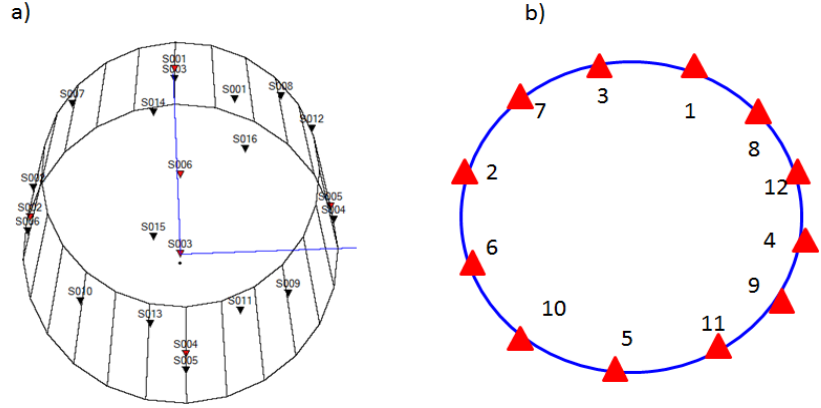

**Figure S.8| Set-up of the sensors (nodes) in our tests in 3D and 2D representations.**

*Acoustic K-chains* To shed more light on the universal pattern of events, we visualize the spatiotemporal evolution of networks in terms of the patterns of links between nodes, where links are defined based on a similarity measure of amplitudes recorded at each sensor at the same time. This is very helpful in understanding the evolutionary phases of Q(t). In addition the trend of $R \equiv Q^{-1}$ is comparable to $<k(t)>$, the average degree of all nodes in the system at a given time. Here $k(t)$ is the degree of each node, defined as the number of links connected to that node at as a function of time and $<...>$ represents the average of all nodes. We visualize the spatial evolution of the degree $k_i$ of the $i^{th}$ node (number of connected links to the node), using polar coordinates $(r_i, \theta_i)_{i=1,...,Nodes}$ where $r_i = k_i$ and $\theta_i$ indicates the position of the node which is fixed on the outer circumference of the cylindrical sample. We refer to these configurations as "*K-chains*", and the normal vector of the K-chains at each node indicates the local direction of increasing/decreasing $k_i$ with time. We evaluate the variation of $r_i = k_i$ at each spatial position (i.e., each node) while we consider the temporal evolution of each single event (Fig.S9). To analyze the spatiotemporal evolution of *K*-chains, chains are simplified by mapping them on the "*spin chain*" where for each node we assign $s_i = sign(\frac{\partial k_i}{\partial t})$ and then $s_i = \pm 1$ (Fig.S9-S11). With this mapping, each node in a given time-step acquires one of the states ($\uparrow$ or $\downarrow$).

In a fully "saturated" state (all arrows direction are up or all are down; all nodes are of identical sign) and the *flipping* of nodes are represented with negative $s_i$ indicated by inward-pointing normal vectors (Fig. S11- Ghaffari et al 2016). This damaged node acts similar to a kink, separating zones with up and down nodes. In Fig.S11 ,we have shown that creation and annihilation of kinks-antikinks (i.e., pairing mechanics) govern the transitions between main phases of Q(t) (or $<K(t)>$). The corresponding frequency-time domain representation of the signals features these (topological) defects as distortions to line-shapes in space-time plots (Fig.S.12b) [8].

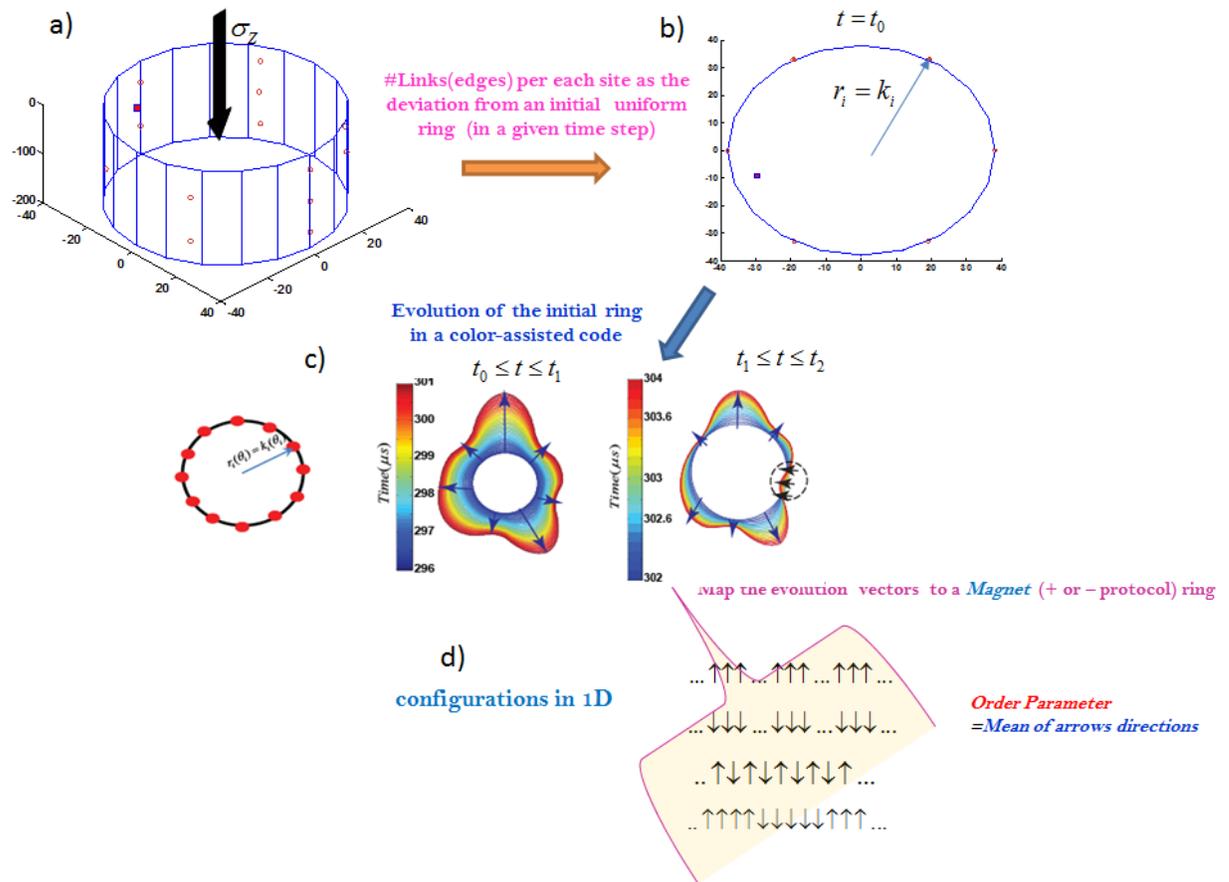

**Figure S.9|** (a-d) Steps to map the K-chains to fictitious spins.

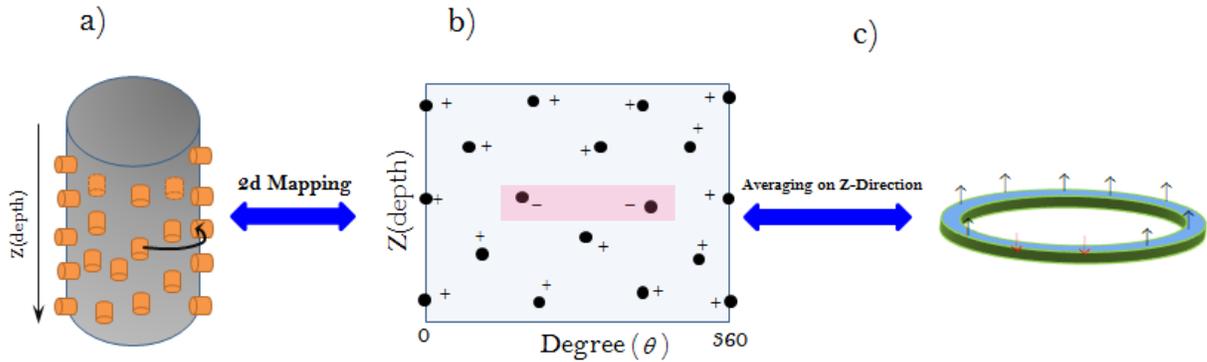

**Figure S.10| mapping a multi-array ultrasound sensors set-up around a cylindrical sample on a 2d and 1d structures (ring-like).** The +/- signs are schematic representation of K-chain's vectors(states)

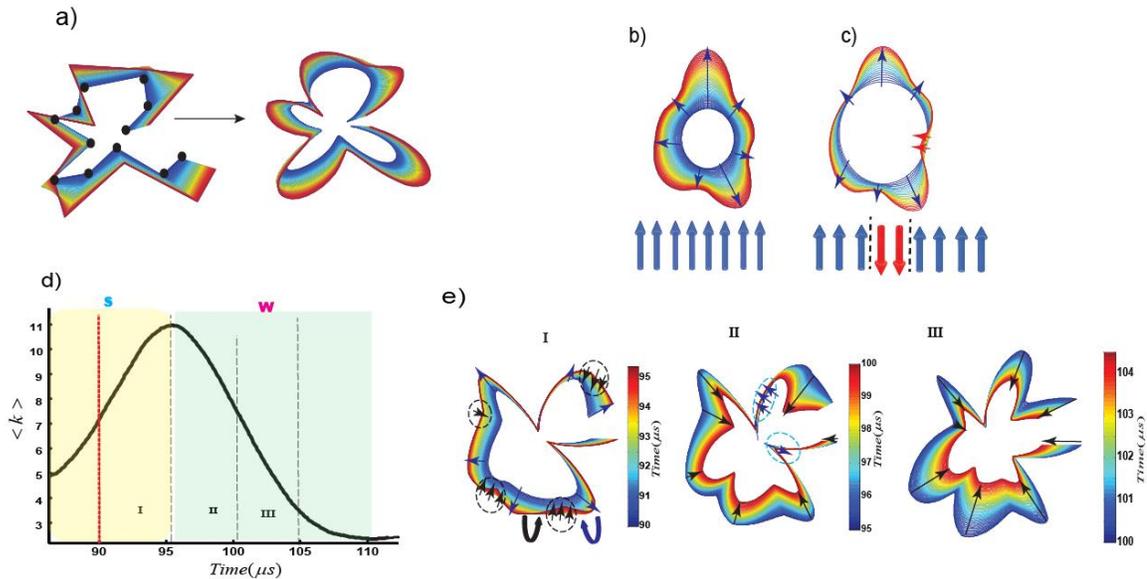

**Figure S.11| K-chains mechanism. a) The original K-chain with 12 nodes representing the 12 acoustic sensors mounted around the basalt specimen is smoothed by spline interpolation.** The result is a plot of 300 nodes that is easier to visualize. (b,c) K-chains are mapped to spin-like systems (min.2D) where each node is set in analogy with a "spin". (d) An example of the dry event: the average of node degrees ($<k>$) versus time (e) Snapshots of *K*-chains showing the evolution (in 2D polar coordinate system) at three time intervals. Folding (crumpling) of the chain is the source of non-linearity and fragments the network with formation of pairs of kinks/anti-kinks (thick blue and black arrows in the panel "I"). At the third time interval, the system is in a fully polarized (ordered) state. Transition from fully ordered state (e.g., *S*-phase) to another fully ordered state (e.g., W-phase) is associated with the creation of "*kinks*" as the flipped nodes.

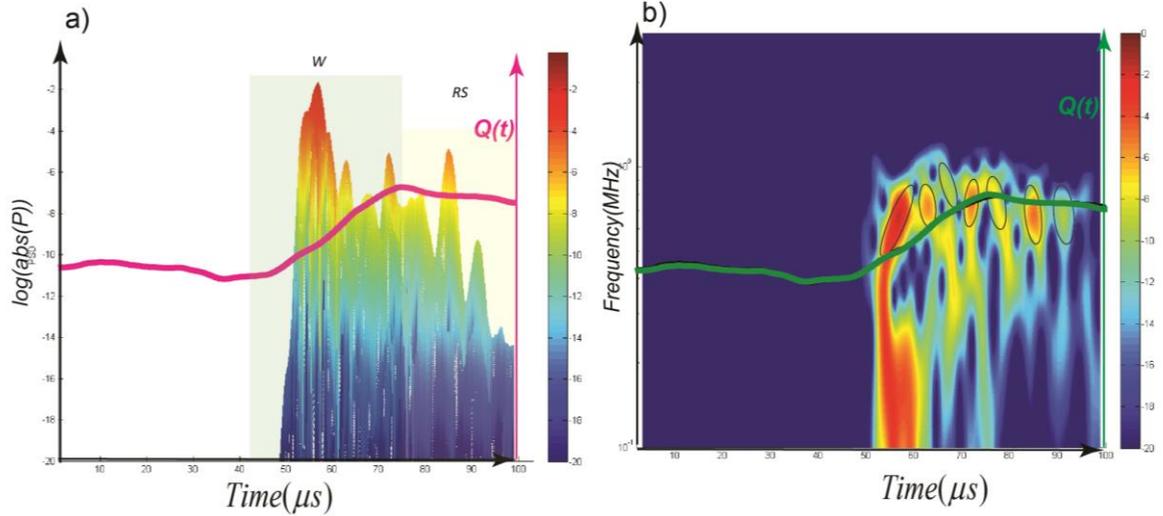

**Figure S.12| Distortion of space-time line-shapes** (elliptical features in semi-logarithmic representation) in **frequency-time plots** of a <u>dry cracking excitation</u>. These diagonal distortions are related to the topological defects [8] as it can be visualized in K-chains.

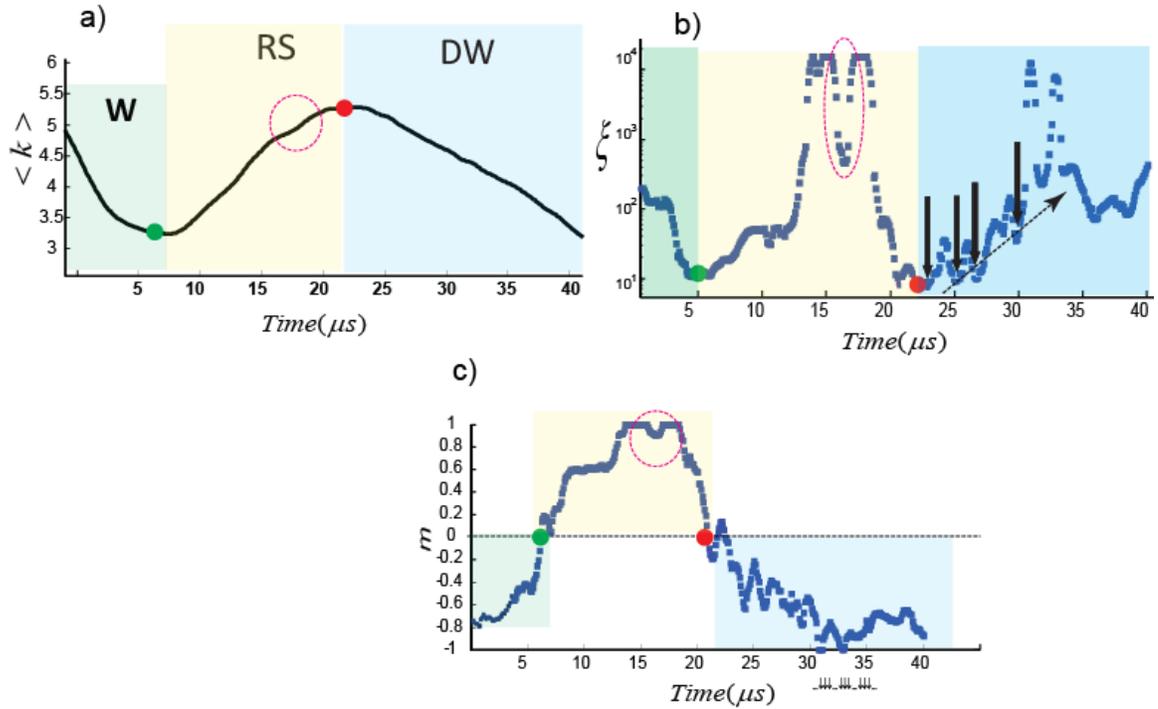

**Figure S.13| Structure of the double weakening phase in Wet-cracking noises. (a,b) the mean degree of all nodes for a hybrid event and the corresponding correlation length of *K*-chains. (c) Highly fluctuating order parameter in DW phase drives the system in to the next degenerate state.** This signature distinguishes the physics of the secondary weakening phase from the W-phase. We assign this fluctuation to discrete process of squeezing-out of the liquid. The frequency of this almost periodic oscillation is around *200-300kHz*. A double peak in RS phase-exhibited in *m*(t) and $\xi(t)$ - induces sequence of stiffening-softening like feature with durations less than 2μs (red dotted-circles).

An example of a hybrid-event is shown in Figure S.14 (also FigS.13). The nodes with the healing-like characteristics elongate in the RS phase, covering a large fraction of the chain's circumference (The dotted red-line around the chain). As an approximation, the mean direction of slip associated with the DW phase might be inferred from the patterns of black-arrows (as defects growth). In this case they expanded in a direction parallel to the elongated healed-like region; the preceding re-strengthened region is re-failed.

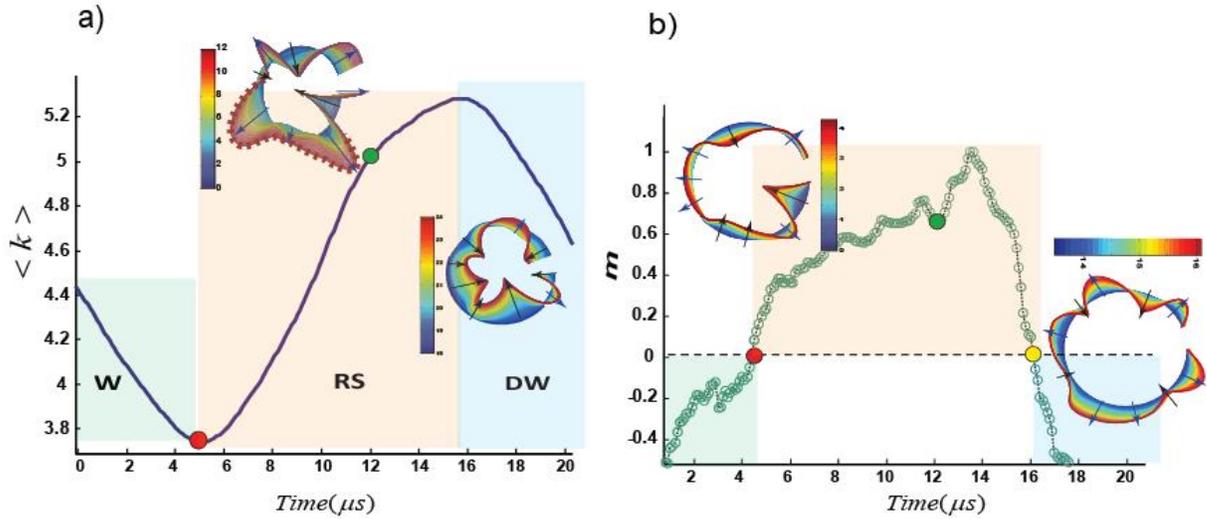

**Figure S.14| K-chains patterns and susceptibility in a wet-cracking noise. (a,b) the average number of links versus time and the real-time order parameter for event #478.** We have shown the temporal cumulative snapshots of K-chains. The blue and black arrows show expansion and contraction of chains, respectively. The dotted red-line around the chain in (a) shows the proliferation of healed nodes, resulting RS phase. The main defect region in DW phase in panel (a) coincides with the red-dotted healed like line.

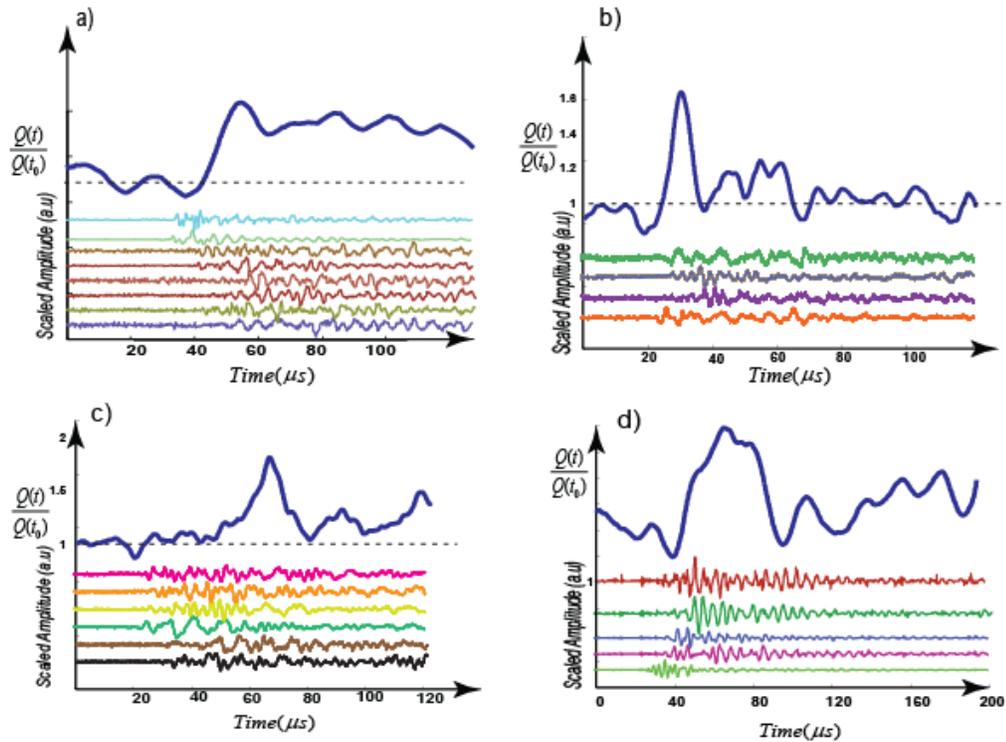

**Figure S.15|(a) Crack-like dry excitation and (b-d) pulse-like microscopic *dry-events* with corresponding multi array-acoustic excitation of ultra-sound waves.**